\documentclass{article}

\usepackage{arxiv}

\usepackage[utf8]{inputenc} 
\usepackage[T1]{fontenc}    
\usepackage{hyperref}       
\usepackage{url}            
\usepackage{booktabs}       
\usepackage{amsfonts}       
\usepackage{nicefrac}       
\usepackage{microtype}      
\usepackage{lipsum}
\usepackage{natbib}
\usepackage{eurosym}
\usepackage{booktabs} 
\newcommand{\BID}{\text{BID}}
\newcommand{\BIT}{\text{BIT}}
\newcommand{\BIND}{\text{BIND}}

\usepackage{algorithm, algorithmic}
\providecommand{\keywords}[1]{\textbf{\textit{Keywords---}} #1}

\usepackage{amssymb}
\usepackage{amsthm}
\usepackage{amsmath}

\DeclareMathOperator*{\argmin}{arg\,min}

\newtheorem{theorem}{Theorem}
\usepackage{newtxmath}
\usepackage{setspace} 
\usepackage{lineno,xcolor} 	
    \usepackage{graphicx} 
    \usepackage{float} 
    \usepackage{subfig}
    \usepackage{printlen}
    
    \usepackage{multirow}
    \usepackage{tabularx}
    \usepackage{array}

\title{VAT tax gap prediction: a 2-steps Gradient Boosting approach}

\author{
  Giovanna Tagliaferri 
  \\
  Business Intelligence and Data Warehouse\\
  SOGEI\\
  Rome, Italy 00143 \\
  \texttt{ggtagliaferri@sogei.it} \\
   \And
 Daria Scacciatelli \\
  \\
  SOGEI\\
  Rome, Italy 00143  \\
  \texttt{daria.scacciatelli@gmail.com} \\
   \And
 Pierfrancesco Alaimo Di Loro \\
  Department of Statistics\\
  University of Rome ``La Sapienza''\\
  Rome, Italy  00185\\
  \texttt{pierfrancesco.alaimodiloro@uniroma1.it} \\
}

\begin{document}
\maketitle



\doublespacing

\begin{abstract}
\noindent Tax evasion is the illegal evasion of taxes by individuals, corporations, and trusts. The revenue loss from tax avoidance can undermine the effectiveness and equity of the government policies. A standard measure of tax evasion is the \textit{tax gap}, that can be estimated as the difference between the total amounts of tax theoretically collectable and the total amounts of tax actually collected in a given period.
This paper presents an original contribution to bottom-up approach, based on results from fiscal audits, through the use of Machine Learning. The major disadvantage of bottom-up approaches is represented by selection bias when audited taxpayers are not randomly selected, as in the case of audits performed by the Italian Revenue Agency. 
Our proposal, based on a 2-steps Gradient Boosting model, produces a robust tax gap estimate and, embeds a solution to correct for the selection bias which do not require any assumptions on the underlying data distribution. 
The 2-steps Gradient Boosting approach is used to estimate the Italian Value-added tax (VAT) gap on individual firms on the basis of fiscal and administrative data income tax returns gathered from Tax Administration Data Base, for the fiscal year 2011.
The proposed method significantly boost the performance in predicting with respect to the classical parametric approaches.
\end{abstract}

\setlength\parindent{.45in} \keywords{Gradient Boosting, Selection Bias, Tax Gap, Machine Learning}


\section{Introduction}
Tax evasion is the illegal evasion of taxes by individuals, corporations, and trusts. The revenue loss from tax avoidance can undermine the effectiveness and equity of the government policies. It represents one of the main problems in modern economies, where the government budget is constantly under control \citep{santoro2010evasione}.  A standard measure of tax evasion is the \textit{tax gap}, that can be estimated as the difference between the total amounts of tax theoretically collectable and the total amounts of tax actually collected in a given period.
Tax gap estimations can be divided into two broad methodological approaches: macro and micro. Methodologies based on a macro perspective (top-down) usually employ macroeconomic indicators or national accounts data. Methodologies based on a micro perspective (bottom-up) employ more specific or individual data derived from administrative tax records provided by internal fiscal agencies, or audit data\footnote{Data derived from ad-hoc tax assessments/controls on the taxpayer} (\cite{pisaniBott}; \cite{dangerfield1992top}; \cite{OECD2017}). 
The bottom-up perspective is used to derive a more robust estimate of single components of the tax gap related to different taxpayers. 

Several tax gap and tax evasion studies and analyses have been conducted on Italy. In particular, over the last few years, the Italian Revenue Agency (IRA) and Sogei have produced some preliminary estimates of the tax gap related to self-employed taxpayers and small firms, through the bottom-up approach based on the results of fiscal audits \cite{taxpap_0073}.
The auditing activity of the Italian Revenue Agency is not based on random audits but, on the contrary, it relies on the so-called risk-based audits: taxpayers selection is not random but driven by risk assessments performed by tax-auditors.
Hence, as of now, the tax evasion estimates are based on the Tobit-like model known as the "Heckman model" in order to correct for the non-random nature of the available sample (\cite{tobin1952survey}; \cite{amemiya1986advanced}; \cite{heckman1979sample}).

The aim of this paper is to present an alternative estimation technique of the tax gap based on the bottom-up perspective of fiscal audits (\cite{braiotta2015tax}; \cite{kumar2015minimising}).
The analysis is focused on the estimation of the Italian Value-added tax (VAT) gap on individual firms\footnote{individuals liable for tax on income as self-employees  persons and  small  individual  companies (ownership, board of directors and management are totally controlled by one person).} by integrating data from different sources as Tax Register (Anagrafe Tributaria) and Tax Audits Database. The Tax Register contains information about Personal Income Tax (PIT), Value Added Tax (VAT) and Regional Tax on Productive Activities (IRAP) from the filed tax returns, namely "Declared Income Tax Base" (BID); Audits Database contains, for a small non-random sample of taxpayers, a correction of the declared taxable base: the "Potential Tax Base" (BIT). 
We want to produce reliable estimates for the potential VAT turnover for the whole population of taxpayers. In practice, exploiting the information collected on the audited taxpayers, we should be able to infer the potential tax base on the non-audited ones and obtain an estimate of the complete Italian Value-added tax (VAT) gap on individual firms.

The main issue of this analysis is that just a small subset of taxpayers is audited (around $2\%$) and they are not randomly selected \citep{d2016general}; this induces a \textit{selection bias} on the observed response variable which invalidate application of standard statistical methods (\cite{sarndal2003model}; \cite{sarndal2005estimation}). More details on the Italian tax system and the operational audits activity are presented in Section \ref{ItTaxSys}.

The focus of this papers stands in presenting a \textit{machine learning} method based on the combination of the Gradient Boosting algorithm and the 2-steps approach proposed by \cite{zadrozny2004learning}. 
The adoption of machine learning algorithms is driven by the necessity to move beyond linearity and exploit their ability to extract information from very large sets of data (both in terms of units and variables) without any strong assumption on their distribution.
In the domain of this paper, the Gradient Boosting algorithm, introduced by \cite{friedman2001greedy}, is chosen over its alternatives. It is a very robust ensemble \textit{learner}, able to deal with data of any size (if adequate computational power is available) and nature. More methodological details are presented in Section \ref{Meth}.
Afterwards, in Section \ref{App}, the proposed methodology is applied to estimate the potential VAT turnover for a representative sample of Italian firms, for the fiscal year $2011$. Results are compared with the ones obtained using the Heckman model, leading to the conclusion that the proposed method produces more robust tax gap estimates, boosting sensibly the predictive performances with respect to the classical Heckman parametric approach (\cite{heckman1976common}; \cite{heckman1979sample}).
Finally, in Section \ref{PredProp}, the propensities to evade VAT tax (the ratio between Value-added tax gap and the declared part) are estimated for not-audited taxpayers. These propensities can be used for detecting high risk taxpayers and drive for future tax audit selection.

\section{Italian Taxation System}
\label{ItTaxSys}
\begin{flushright}
{ \textit{"\textbf{Every} person shall contribute to public expenditure in accordance with his/her tax-payer
capacity. The taxation system shall be based on criteria of progression."}
{\\\footnotesize \textit{Const. Art. 53, Section I, Political rights and duties}}}
\end{flushright} 

The taxation system in Italy is based on three fundamental principles, established by Italian Constitution in the Rights and Duties of Private Citizens section.

\begin{itemize}
\item  Universality of taxation: all citizens must contribute to public expenses through the payment of the taxes. These are aimed to finance the operation of the state machine and are reflect in terms of performance and services for citizens. Those who are below a minimum income are exempt from the tax obligation and can also take advantage of all services, by virtue of the principle of economic and social solidarity.
\item  Ability-to-pay taxation: is a progressive taxation principle that maintains that taxes should be levied according to a taxpayer's ability to pay. This progressive taxation approach places an increased tax burden on individuals, partnerships, companies, corporations, trusts, and certain estates with higher incomes.
\item Criteria of progression: the payment of taxes by citizens varies proportionally with respect to the potential tax base. This means that everyone pays taxes based on their economic possibility with a contribution that grows as income increases.
\end{itemize}
The most important sources of tax revenues in Italy are Personal Income Tax (PIT), Value Added Tax (VAT), Corporate Income Tax (CIT) and Regional Tax on Productive Activities (known as IRAP). 
\\The individual personal tax base (so called IRPEF) for employees, retired workers and self-employed  is the most important tax in the Italian system in terms of revenue (equal to almost $40\%$ of total tax revenue); resident tax subjects are subject to Italian personal (or national) income taxes on their income wherever produced (according to the so-called "world principle"). The study of PIT evasion and its gap is addressed in \cite{braiotta2015tax}.
\\ The regional tax on productive activities affects wealth at the stage of its production and not at that of its perception (such as income taxes) nor at its consumption (as we will see for VAT). The subjects involved are the freelancers, institutions and companies.
\\ The Value Added Tax is a general tax on consumption that is applied in Italy and in all other European Union states. The European Commission states: <<\textit{it serves to tax the consumption of goods and assets services. It is applied to all commercial activities involving the production and distribution of goods and the provision of services}>>. This kind of tax in not progressive and weighs completely on the final consumer, while for the taxable person (the entrepreneur and the self-employed) it remains neutral. In Italy, it provides about $25\%$ of the annual tax revenue. From a theoretical point of view, its revenue is not affected by the length of production chain and distribution since it is collected fractionally. This aspect assures the neutrality to the degree of vertical integration of the production process and to the steps that goods and services follow before being purchased by final consumers \citep{d2014asymmetries}.

The Italian Revenue Agency (IRA) verifies the level of spontaneous fulfilment of tax obligations by taxpayers. The IRA activity is to evaluate, with different methods and different timing, the correspondence between what is declared and what is actually due \citep{d2016general}.
The IRA has up to five years to audit a taxpayer report after it is filed; for instance, at December 31 2017 has been completed audit processes for tax declarations presented in 2012 (fiscal year 2011).
In general, the audited taxpayers are not selected randomly but they are identified on the basis of predetermined selection criteria. The underlying assumption is that non-compliant taxpayers have a different behavior from the compliant ones. Non-random audits are quite expensive and time-consuming and, for this reason, they usually involve a very small number of taxpayers (less than $10\%$ of the taxpayer population). A robust methodology to estimate and predict the undeclared part for non-audited taxpayers could better support decision for policy-makers and contrast tax avoidance.

\section{Modeling approach: purpose, objective, and methodology}
\label{Meth}
Tax non compliance is a phenomenon that not only directly affects revenue sufficiency through reduced tax revenues, but also impacts on income distribution and equity, limiting development and sustainable economic growth.
Indeed, the amount of tax revenue actually collected by the state (declared tax base, $\BID$) is generally lower than the true potential tax base ($\BIT$) because of the presence of an undeclared part ($\BIND$).
This undeclared part can be straightforwardly derived from $\BIT$ and $\BID$ using the main formula of the gap computation:
\begin{equation*}
    \BIND = \BIT-\BID .
\end{equation*}

This work is focused on the estimation of the Italian Value-added tax (VAT) gap on individual firms by integrating data from different sources as Tax Register (Anagrafe Tributaria) and Tax Audits Database.
While data on individual income taxes $\left\lbrace\BID_i\right\rbrace_{i=1}^N$ are available for whole population of taxpayers $\mathcal{P}$, the VAT \textit{Undeclared} tax base is generally available only for a small non-random sample of audited taxpayers.
The proposed bottom-up approach, based on the results of audit activities, provides us with the $\BIT$ detected on a sample $s$ of \textit{audited} taxpayers. The tax returns of these subjects are selected on the basis of unknown criteria established by the Director of the Revenue Agency \citep{d2016general}.
This means that audit selection mechanism depends on different covariates that are likely to be correlated with the response variable and therefore, marginally, non-negligible in relation to this one.
Therefore, any estimation procedure based this non-random sample could be affected by a significant  \textit{selection bias} which can invalidate the inference from the sample to the whole population of taxpayers  \citep{sarndal2005estimation}.

For instance, let us denote our variable of interest $Y$ and let it be observed only on a non random sample of units $s$ selected according to a sample design $\mathcal{S}$; the expected value for each unit would be:
\begin{equation}
    \mathbb{E}[Y_i] = \mathbb{E}[Y_i|i\in s]P(i\in s)+\mathbb{E}[Y_i|i\in s^c]P(i\in s^c)
\end{equation}
If the sample design $\mathcal{S}$ is not independent from the outcome variable, then:
\begin{equation}
    \mathbb{E}\left[Y_i|i\in s\right] \neq \mathbb{E}\left[Y_i|i\in s^c\right]
\end{equation}
and it is not possible to directly get any estimate for $\mathbb{E}\left[Y_i|i\in s^c\right]$ using only the observed outcomes. Hence, it is not possible to get any estimate for the potential tax base of not-audited units by using only information on the audited ones. 
Furthermore, due to the confidential nature of the selection criteria, the probability of selection of each unit in the sample $\pi_i$ are not available a-priori and therefore cannot be used to correct for the non-rappresentativeness of the selected sample.

These observations are key in identifying the appropriateness of the chosen model assumptions, by providing with methods of detecting and correcting sample selection bias.

\subsection{The Heckman model: a brief review}
\label{heckModel}
The Heckman two-stage estimation procedure is an econometric tool that allows analysts to take into account the probability of selection \citep{heckman1979sample}. The Italian Revenue Agency has already used the \textit{Heckman Model} to estimate tax gaps and other fiscal authorities are also considering this approach.
This model assumes a direct effect of the outcome variable on the selection probability of each unit. 
The outcome $\left\lbrace y_i\right\rbrace_{i=1}^N$ is modeled through a bivariate latent process $\left\lbrace (z_{i1}, z_{i2})\right\rbrace_{i=1}^N$. The first latent component $z_{i1}$ is the actual variable of interest, while the $z_{i2}$ represents the propensity to be selected of unit $i$. Given two set of covariates $\left\lbrace\mathbf{x}_{i1}\right\rbrace_{i=1}^N$ and $\left\lbrace\mathbf{x}_{i2}\right\rbrace_{i=1}^N$, the two latent components are expressed as:
\begin{equation*}
    \begin{aligned}
    z_{i1} = \mathbf{x}_{i1}^T\boldsymbol{\beta_1} + u_{i1}\\
    z_{i2} = \mathbf{x}_{i2}^T\boldsymbol{\beta_2} + u_{i2}
    \end{aligned}
\end{equation*}
and their relationship with the outcome of interest is the following:
\begin{equation*}
y_i=
    \begin{cases}
    z_{i1}\qquad  &z_{i2}>0\\
    \text{Unknown}\qquad  &z_{i2}\leq 0
    \end{cases}
\end{equation*}
If the two latent components are not independent, then the selection is not random with respect to the variable of interest. In particular, Heckman assumes that $u_{i1}$ and $u_{i1}$ have a bivariate Normal distribution:
\begin{equation*}
    \begin{bmatrix}u_{i1}\\
    u_{i2}\end{bmatrix}\sim\mathcal{N}_2\left(\begin{bmatrix}0\\
    0\end{bmatrix}, \begin{bmatrix}\sigma^2_1 & \sigma_{12}\\
    \sigma_{21} & \sigma^2_2\end{bmatrix}\right)\qquad \forall i=1,\dots, N
\end{equation*}
An expression of the likelihood of this model can be found in \cite{amemiya1986advanced}. However, a full-likelihood based inference was initially discarded in \cite{heckman1979sample} due to the too long computing time it would have required. Heckman's proposal is based on a \textit{limited information maximum likelihood}, where the sample selection is characterised as a special case of omitted variable problem if OLS were used on the observed sample $\left\lbrace y_i\right\rbrace_{z_{i2}>0}$. In this context, the omitted variable is the so-called \textit{Inverse Mills-Ratio}:
\begin{equation*}
    \lambda(\mathbf{x}_{i2}^T\boldsymbol{\beta_2}/\sigma_2)=\frac{\Phi(-\mathbf{x}_{i2}^T\boldsymbol{\beta_2}/\sigma_2)}{1-\Phi(-\mathbf{x}_{i2}^T\boldsymbol{\beta_2}/\sigma_2)},
\end{equation*}
that can be estimated by way of a \textit{Probit} model on:
\begin{equation*}
s_i=
    \begin{cases}
    1\qquad  &z_{i2}>0\\
    0\qquad  &z_{i2}\leq 0
    \end{cases}.
\end{equation*}
We can then get an unbiased estimation of the model on the $\left\lbrace y_i\right\rbrace_{z_{i2}>0}$ by using the following regression:
\begin{equation}
    y_i=\mathbf{x}_{i1}^T\boldsymbol{\beta_1}+\beta_\lambda\lambda(\mathbf{x}_{i2}^T\boldsymbol{\hat{\beta_2}}/\hat{\sigma_2})+\epsilon_i
    \label{omVarOls}
\end{equation}
where $\beta_\lambda=\frac{\sigma_{12}}{\sigma2}$ and $\epsilon_i\overset{iid}{\sim}\mathcal{N}(0, \sigma^2)$.
Sign and magnitude of $\beta_\lambda$ summarize direction and intensity of the relationship between outcome variable and selection process.
This estimation method can be proved to be consistent as long as the normality of $u_2$ hold and it is currently the standard way to obtain final estimates for the Heckman model.

However, even if the model proposed by Heckman looks elegant and can provide an effective solution in a lot of real world applications, it is not devoid of criticism \citep{puhani2000heckman}. For instance, it is generally not possible to distinguish a priori which covariates should affect the selection process and which the outcome variable. In these cases, $\mathbf{X}_1$ and $\mathbf{X}_2$ may have a large set of variables in common or even be \textit{identical}, making two main complications to arise. First of all, Equation \ref{omVarOls} is only identified through the non-linearity of the \textit{Inverse Mills Ratio} and, since $\lambda(\cdot)$ is an approximately linear function over a wide range of its arguments, collinearity problems are likely to affect stability and robustness of the estimates. Furthermore, if the selection depends on covariates that also affect the outcome, then the observed sample $\left\lbrace(y_i, \mathbf{x}_{i1}\right\rbrace_{z_{i2}>0}$ will not be representative of the whole population with respect to the covariates in common. If the relationship between these covariates and the response variable is perfectly linear this would not be an issue, but when linearity is just an approximation this may lead to sensibly wrong estimation of the corresponding slope coefficients. This happens because part of the range of these predictors is only sparsely observed and errors on this portion of the space are consequently under-weighted.

Moreover, one can discuss whether the hypotheses of the Heckman model actually suits our application. As a matter of fact, the \textit{Italian Revenue Agency} selects the sample of taxpayers to audit only through the set of available covariates $\mathbf{X}$, trying to include those taxpayers that are expected to have a larger $\BIND$ given $\mathbf{X}$. This means that the relationship between selection process and outcome is not direct, but indirect and in particular driven by the same set of predictors. This exacerbates the issues discussed above and emphasise the need for alternative solutions.

\subsection{The 2-steps Gradient Boosting approach}
\label{2steps}
The approach considered in this paper is based on the \textit{2-steps} procedure introduced in \cite{zadrozny2004learning}, which relies on re-weighting observations according to their estimated selection probability. This method assumes that there is an \textit{indirect} effect of the outcome variable on the selection scheme: the outcome variable affects the sample selection only through the available covariates. In particular, our proposal exploits the Gradient Boosting algorithm (described in Section \ref{GB}) as a strong learner since it allows to detect non-linear relationships between covariates and outcome and to relax some strong distributional assumptions hardly matched by real data.
Comparative merits of the proposed approach and the Heckman model are investigated in the toy example of Appendix \ref{Toy}

The considered estimation scheme is viable when the complete population list $i=1,\dots,N$ and a common set of covariates $\left\lbrace X_i\right\rbrace_{i=1}^N$ is available on each unit. The following hypotheses need to hold.
\begin{itemize}
\item The probability to be included in the sample for unit $i$ depends only on its covariates $X_i$.
\item The response variable of unit $i$, $Y_i$, is conditionally independent from the sampling design given the covariates $X_i$:
\begin{equation*}
P(Y_i|X_i,i\in s)=P(Y_i|X_i)\quad \forall\, i\in\lbrace{1,\dots,N}\rbrace
\end{equation*}
\end{itemize}
The second assumption allows us to transfer all the dependence of the outcome variable on the sample design to the set of covariates:
\begin{equation}
\mathbb{E}[Y_i|X_i,i\in s]=\mathbb{E}[Y_i|X_i]\quad \forall\, i\in\lbrace{1,\dots,N}\rbrace
\label{expVal}
\end{equation}

Theoretically, the last assumption (see Equation \ref{expVal}) allow to fit the model considering only the selected units.
However, there is still an issue of non-rapresentativeness of the sample with respect to the whole population. Being units in the sample different from units out of the sample both in terms of response and covariates, an estimation based only on the selected sample would rely on a certain dose of extrapolation. The resulting estimates would favor the fit on over-represented units and disregard the fit on under-represented ones.
The 2-steps solution from \cite{zadrozny2004learning} is adopted to correct this kind of bias using a Horvitz-Thompson style estimation, directly derived from the most basic survey sampling theory \citep{sarndal2003model}, exploiting all the auxiliary information available on the population.
The proposed approach consists of the subsequent application of two predictive models on the available data.
\begin{enumerate}
\item \textbf{Classification model}. The first learner is a classifier that is trained on the whole sample and targets the binary variable \textit{selected in the sample} and \textit{not selected in the sample}. It finds and reveals important patterns and regularities in the selection mechanism of units, so that the selection probabilities can be estimated according to the auxiliary information included in the covariates $X$. Estimates of the selection probabilities $\left\lbrace\hat{\pi}_i\right\rbrace_{i=1}^N$ can be produced for the whole population, providing an approximation to the first order inclusion probabilities:
\begin{equation}
\hat{\pi}_i\approx \pi_i=P(i\in s|X_i), \qquad i=1,...,N
\label{incProb}
\end{equation}
\item \textbf{Regression model}. The second learner is a regression model, trained only on the units for which the dependent variable is observed, that targets the \textit{response} variable. In this second step, it is now possible to incorporate the inclusion probability resulting from Equation \ref{incProb} as individual weights in order to correct for the non-representativity of the selected units\footnote{Even if not discussed in this paper, it is also possible to add the estimated inclusion probabilities as an additional covariate and use an Heckman-style correction}. These can be used to produce the inverse weights defined as:
\begin{equation}
\nu_i=\frac{P(i\in s)}{\hat{\pi}_i}\propto\frac{1}{\hat{\pi}_i},\qquad i\in s.
\label{pesi}
\end{equation}
where $P(i \in s)$ is the probability to be selected notwithstanding the set of covariates.
Formula \ref{pesi} stems from the \textit{Bias Correction Theorem}, which states the following.
\begin{theorem}
For all distributions $D$, for all classifiers $h$ and for any loss function $l(h(X),y)$, if we assume that $P(s|X,y)=P(s|X)$ (that is, $s$ and $y$ are independent given $X$), then:
\begin{equation*}
    \mathbb{E}_{X,y\sim D}\left[l(h(X),y)\right]=\mathbb{E}_{X,y\sim \tilde{D}}\left[l(h(X),y)|s=1\right],
\end{equation*}
where $\tilde{D}\equiv P(s=1)\frac{D(X,y,s)}{P(s=1|X)}$
\end{theorem}
In practice, weighting each input observation $\left\lbrace Y_i, X_i\right\rbrace_{i\in s}$ proportionally to the inverse of their selection probability, we reduce the importance of units already over-represented in the sample while increasing the importance of under-represented ones.
\end{enumerate}
Solutions of this kind are very common in the correction for bias deriving from non-negligible sampling designs \cite{sarndal2003model}, for instance when incorporating the response probability to correct for the non-response bias (\cite{bethlehem1988reduction}; \cite{alho1990adjusting}) or in the case of the inverse probability of treatment weighting (IWTP) (\cite{hirano2003efficient}; \cite{austin2015moving}).
Naturally, all these methods rely on the accuracy of prediction of both models: the classifier in the first step and the predictive model (either classifier or regressive) in the second one. The models choice in the two steps depends on the nature of the problem and the researcher knowledge and sensibility plays an important role. 
This work proposes the adoption of multiple \textit{Classification and Regression Trees} (CART) ensambled through the Gradient Boosting algorithm for both the steps \citep{breiman1984classification}.

\subsection{The Gradient Boosting algorithm}
\label{GB}
The \textit{Gradient Boosting} is a very powerful algorithm that allow to build predictive models for both the classification and regression tasks. It is an ensemble algorithm that relies on the concept of boosting, which is a technique for reducing bias and variance in supervised learning, firstly introduced in the seminal paper of \cite{schapire1990strength}.
The \textit{Gradient} in front of the term \textit{Boosting} refers to a very flexible formulation of the boosting, firstly proposed by \cite{friedman2001greedy}. This particular version exploits the \textit{Gradient Descent} in order to fasten the optimization procedure on the loss function.
This has been chosen among a set of Machine Learning algorithms for both steps because of its desirable combination of reduced computation burden and good performances in either tasks.

Let us consider the usual set of covariates $X=\lbrace x_1,...,x_N \rbrace\in\mathcal{X}$ and the response variable $Y\in\mathcal{Y}$. The final aim of any supervised learning algorithm is to train itself on a set of data $\left\lbrace X_i,Y_i \right\rbrace_{i=1}^N$ whose covariates and response variables are known and then produce an approximation $F^*(x)$ to the function $F(X):\mathcal{X}\rightarrow\mathcal{Y}$ that generally relates $X$ and the expected value of $Y|X$. The approximation is obtained in such a way that the expected value of a pre-specified loss function $L(Y, F(X))$ is minimized with respect to the joint distribution of all the pairs $(X, Y)$ in the set of data.
In practice, the algorithm learns from the examples provided to it in the form of a training set and it looks for that approximation $F^*$ such that:
\begin{equation*}
F^* = \argmin_{F} E_{Y,X} L(Y, F(X))=
\argmin_{F} E_{X}[E_y (L(Y, F(X)))|X].
\end{equation*}
The choice of the loss function depends on the nature of the problem and of the outcome variable. For instance, in the case of the regression task, the usually adopted loss function $L(Y, F(X))$ is the \textit{Mean Squared Error}.
The peculiarity of the boosting procedure is that it approximates $F(X)$ using a function of the form:
\begin{equation*}
F^*(X) = \sum_{m=0}^M \beta_m h_m(X)
\end{equation*}
where $h_m(X)$ are functions known as \textit{Base Learners} and $\left\lbrace\beta_m\right\rbrace_0^M $ are real coefficients. The base learners are functions of $X$ derived from another, simple, learning algorithm and the $\beta$'s are expansion coefficients used to combine the base learners outcomes.
Either the base learners and the expansion coefficients are estimated using the data from the train set using a \textit{forward-stagewise} procedure.
As any recursive algorithm, it starts from an initial guess $F_0(X)$ and then the new set of coefficients and learner are derived as:
\begin{equation}
(\beta_m,h_m)=\argmin_{\beta,h}\sum_{i=1}^N L(Y_i, F_{m-1}(X_i)+\beta h(X_i))\quad \forall \;m=1,...,M
\label{gb1}
\end{equation}
and
\begin{equation*}
F_m(X) = F_{m-1}(X)+\beta_m h_m(X_i,a))
\end{equation*}

\begin{algorithm}
\begin{doublespace}
\begin{algorithmic}
\STATE {$F_0(X)= \argmin_{\rho,\beta} \sum_{i=1}^N L(y_i,\beta)$}
\FOR  {$m=1$ up to $M$}
        \STATE $\tilde{Y}_{i,m} =- \left[ \dfrac{\partial L(Y_i, F(X_i))}{\partial F(X_i)} \right] _{F(X)=F_{m-1}(X)},\,\forall\; i\in \left\lbrace1,...,N\right\rbrace$;
		\STATE $h_m=\argmin_{h}\sum_{i=1}^NL \left(\tilde{Y}_{i,m}-h(X_i)\right)$
		\STATE $\beta_m=\argmin_{\beta}\sum_{i=1}^N L(Y_i, F_{m-1}(X_i)+\beta h_m(X_i))$
		\STATE $F_m(\mathbf{x})= F_{m-1}(X)+\lambda\beta_m h_m(X_i))
$
\ENDFOR
\end{algorithmic}
\caption{Gradient Boosting pseudo-code example}
\label{gbAlgo}
\end{doublespace}
\end{algorithm}

Unfortunately, choosing the best pair $(\beta_m,h_m)$  at each step for an arbitrary loss function is a computationally infeasible optimization problem in general. This is where the gradient descent plays a key role, leading to the \textit{Gradient Boosting} algorithm.
It solves the optimization problem in Equation \ref{gb1} through an approximation that is legit whenever the loss function $L(Y, X)$ is differentiable. 
At each step $m=1,\dots,M$, the base learner $h(X)$ is chosen according to the best fit on the \textit{pseudo-residuals} $\left\lbrace\tilde{Y}_{i,m}\right\rbrace_{i=1}^N$, deriving from the previous step:
\begin{equation*}
h_m = \argmin_{h}\sum_{i=1}^NL \left(\tilde{Y}_{i,m}-h(X_i)\right),
\end{equation*}
where:
\begin{equation*}
\tilde{Y}_{i,m} = Y_i-\rho F_{m-1}(X),\qquad i=1,\dots,N,\; \rho\in\mathbb{R}^+.
\end{equation*}
The pseudo-residual values $\left\lbrace\tilde{Y}_{i,m}\right\rbrace_{i=1}^N$ play the role of the gradient, driving the optimization procedure towards the right direction step after step.
In this simplified framework, given the base learner $h_m(X)$, the best value $\beta$ for $\beta_m$ can be obtained as:
\begin{equation*}
\beta_m=\argmin_{\beta}\sum_{i=1}^N L(Y_i, F_{m-1}(X_i)+\beta h(X_i)).
\end{equation*}
A very common modification to the standard gradient boosting algorithm is the addition of a \textit{shrinkage parameter} $\lambda$, which modifies the update rule in the following way:
\begin{equation*}
F_m(X) = F_{m-1}(X)+\lambda \beta_m h_m(X_i,a)),\qquad m=1,\dots,M,\;\lambda\in [0,1].
\end{equation*}
This parameter controls the learning rate of the algorithm and allows for the regularization of the procedure  \citep{efron2016computer}.
The whole algorithm is resumed in the pseudo-code Algorithm \ref{gbAlgo}.

The most common version of the Gradient Boosting uses fixed-size CART (usually small, with low number of branches and/or splits) as base learners, whose predictive ability is strongly enhanced by their boosting combination \citep{efron2016computer}.
Either the shrinkage parameter and the parameters that define each single random tree (number of splits, number of branches, etc.) are not estimated during the procedure. In the Machine Learning context they are known as \textit{tuning parameters} and they need to be chosen in advance and stay fixed. Typically, they are selected via searching procedure based on the \textit{cross-validation} in order to avoid over-fitting \citep{friedman2001elements}.

\subsection{Estimation of uncertainty}
\label{Unc}
The main issue with machine learning algorithm is that they provide point estimates but they cannot rely on any modeling assumption in order to derive interval estimates. However, it is possible to address this weakness resorting to a bootstrap approach \citep{efron2003second} as it is proposed in \cite{heskes1997practical}\footnote{Not to be confused with the \textit{stochastic gradient boosting} method that may be used in the optimization process of the predictive algorithm}.

\begin{figure}
    \centering
    \includegraphics{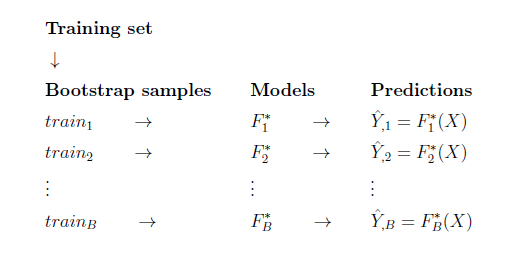}
    \caption{Bootstrap procedure scheme.}
    \label{Boots}
\end{figure}

This method consists of deriving $B$ different bootstrap samples from the original training set, in order to get $B$ different samples approximately distributed according the joint distribution of $Y$ and $X$ of the considered training set.
The idea is to fit the 2-steps GB on the $B$ different samples from the original training set, which will provide a different approximation for the function relating covariates and the expected value of the response variable $\left\lbrace F^*_j(X)\right\rbrace_{j=1}^B$. 
Using each of these functions, it is possible to produce $B$ set of predictions $\left\lbrace\hat{Y}_{\cdot j}\right\rbrace_{j=1}^B$ for all the $Y_i$'s in the training set: $B$ different predictions for each unit $i=1,\dots,N$. The procedure is outlined in Figure \ref{Boots}.

In such a way, a sort of empirical posterior distribution for the prediction $\hat{Y}_i$ of each $Y_i$ is obtained and, through these, it is possible to derive interval estimates in whatever way it is preferable. For instance, it may be considered picking the $\alpha/2$ and $1-\alpha/2$ empirical quantiles in order to obtain a $(1-\alpha)\%$ equal tail posterior interval for the prediction.


\section{Estimation of the Italian Value-Added Tax (VAT) gap}
\label{App}

The considered dataset is composed of all the Italian individual firms included in Tax Register for the fiscal period 2007-2014. In according with the Italian fiscal law,  these individuals are liable for tax on income as self-employees  persons and  small  individual  companies (ownership, board of directors and management are totally controlled by one person).

An unavoidable delay occurs between the availability of the audit data and the fiscal year of reference. 
First, there is a lag between the fiscal (audited) year and the year in which the control is performed. Second, due to the characteristics of the auditing process performed by the IRA, there is the need to wait (on average) two years after the year of the audit to have final data.
On the other hand, one tax year can be audited only until a maximum of $5$ years after the corresponding tax return has been completed. In view of the above, estimates on a tax year are available within six to seven years from the fiscal year of reference and, for these reasons, the following analysis will be driven on data referred to the year $2011$.
The population of interest consists of $N=2.3$ millions ($2'293'937$) of individual firms, where only the $0.82\%$ ($18'718$ units) have been audited.
The database includes about $159$ variables coming from the Tax Register that concern various area of information about the owner and its firm: personal data; economic sector of operation; taxable income and tax by type; revenues, expenses incurred, taxable base, gross and net tax; presumptive turnover provided by \textit{Business Sector Studies}. A more detailed list of the available covariates is reported in Table \ref{covars}.
\begin{table}
\setlength{\extrarowheight}{2pt}
\begin{tabular}{ |l|l| }
\hline
\multirow{5}{8em}{Personal data}
    & $\bullet$ Age (individual)\\
    & $\bullet$ Gender (individual) \\
    & $\bullet$ Fiscal regime (individual)\\
    & $\bullet$ Branch of economic activity (firm)\\
    & $\bullet$ Region (firm)\\
\hline
\multirow{3}{8em}{Audit (only audited) $\qquad$ $\qquad$ }
    & $\bullet$ Potential volume of business (BIT) \\
    & $\bullet$ Undeclared volume of business (BIND) \\
    & $\bullet$ Assessment indicator variable \\
\hline
\multirow{6}{8em}{IRAP}
    & $\bullet$ Amortization \\
    & $\bullet$ Operating costs \\
    & $\bullet$ IRAP total revenues \\
    & $\bullet$ IRAP total tax \\
    & $\bullet$ Net production value\\
    & $\bullet$ Revenues \\
\hline
\multirow{12}{8em}{PIT}    

    & $\bullet$ Personal Income Tax \\
    & $\bullet$ Labour cost (amount of) \\
    & $\bullet$ Negative components of income (costs and expenses)  $\qquad$ $\qquad$ $\qquad$ \\
    & $\bullet$ Revenues from activities \\
    & $\bullet$ Gross income \\
    & $\bullet$ Income \\
    & $\bullet$ Total remuneration  \\
    & $\bullet$ Tax deductions  \\
    & $\bullet$ Input/Output Tax  \\
    & $\bullet$ Other incomes\\
    & $\bullet$ Total expenses  \\
    & $\bullet$ Profit  \\
\hline
\multirow{10}{8em}{VAT}    

    & $\bullet$ Operations generating VAT  \\
    & $\bullet$ Value Added Tax \\
    & $\bullet$ Purchases and imports\\
    & $\bullet$ Total VAT credit \\
    & $\bullet$ Volume of business (BID) \\
    & $\bullet$ Input/Output Tax \\
    & $\bullet$ Taxable transactions  \\
\hline
\end{tabular}
\caption{Summary of covariates included in the Tax register organized by category.}
\label{covars}
\end{table}
The analysis was carried out using the open-source software $\texttt{R}$, considering all the most recent features that allow the management of large amounts of data (i.e. \cite{wickham2015dplyr}; \cite{wickham2016package}).
However, the Tax Register Database contains data considered 'sensitive' and subjected to specific processing conditions, that could not be moved in any case to external virtual machines. Therefore, the analysis had be performed on a single computer (Processor: Intel Pentium dual-core E1040; RAM: 4gb), with a particularly limited amount of RAM. This had no effect on the algorithm performance, but did not allow us to consider the whole dataset of individual firms.
Therefore, only a stratified sample of all the units has been considered, where the not-audited units have been under-sampled controlling for the demographic variables, while all the audited units have been included in the analysis.
In particular, three strata have been considered for the selection of non-audited individuals: Fiscal regime, regions and branch of economic activity (ATECO).

It is relevant to point out that the sub-sampling on not-audited taxpayers directly affects only first step of the proposed approach (Section \ref{2steps}) and only indirectly the second one.
Indeed, the estimation of inclusion probability could be less accurate, and affect the selection bias correction.
However, \textit{under-sample of the majority class} (\cite{chawla2009data}; \cite{he2009Imb}) is one of the most common strategies to handle imbalanced data and, given the actual great imbalance between audited and not-audited taxpayers in our population, this sub-sampling choice may even improve the final estimates \citep{more2016survey}.
Finally, the considered sample consisted of $64'207$ individual firms: $45'489$ not-audited taxpayers and $18'718$ audited ones (see Table \ref{subsam}).

\begin{table}
\footnotesize
\centering
\begin{tabular}{lcccc}
\toprule
& \multicolumn{2}{c}{\textbf{Total Population}} & \multicolumn{2}{c}{\textbf{Sub-sample}} \\
\textbf{Fiscal audits} \qquad &  \textbf{Frequency} \qquad &  \textbf{Percentage} \qquad &  \textbf{Frequency} \qquad &  \textbf{Percentage} \\
\midrule
Not-audited \qquad & $2'275'219$  \qquad  &  $99.18\%$ \qquad & $45'489$   \qquad  &  $70.85\%$\\
Audited \qquad & $18'718$  \qquad  &  $0.82\%$ \qquad & $18'718$   \qquad  &  $29.15\%$ \\
\midrule
\qquad & $2'293'937$  \qquad  &  $100\%$ \qquad & $64'207$ \qquad  &  $100\%$ \\
\bottomrule
\end{tabular}
\caption{Total and sampled population of individual firms.}
\label{subsam}
\end{table}

As described previously, this paper is focused on the estimation of the Italian Value-added tax (VAT) gap on individual firms, according to procedure introduced in Section \ref{Meth} and compare its performances to the Heckman parametric model.
We recall that the declared tax base $\left\lbrace\BID_i\right\rbrace_{i=1}^N$ and all the covariates $\left\lbrace X_i\right\rbrace_{i=1}^N$ have been retrieved from the Tax Registry and are available on the whole population. The effective potential tax base $\left\lbrace\BIT_i\right\rbrace_{i\in s}$ is retrieved from the fiscal audits papers and is available only on a non-random sample of units subject to controls.
The undeclared tax base$\left\lbrace\BIND_i\right\rbrace_{i=1}^N$ is derived by subtracting from BIT the declared tax base $\left\lbrace\BID_i\right\rbrace_{i=1}^N$, following the main formula of the gap computation: 
\begin{equation*}
\BIND_i = \BIT_i-\BID_i,\qquad i=1,\dots,N.
\end{equation*}
Moreover, the prediction of the amount of BIND for not-audited taxpayers are used to analyze propensities to evade VAT tax for different subgroups; these propensities can be used to detecting high risk taxpayers and drive for future tax audit selection.

The pre-processing of the data consisted of joining the information derived from all the different sources and dropping unary variables and variables with high percentage (more than $80\%$) of missing values. Further data pre-processing has been considered (standardization, decorrelation, etc.), but did not lead to any performance improvement; this is not suprising since machine learning methods are not affected by the usual data criticism of standard linear modeling such as: multicollinearity, skewness, deviations from Normality assumptions and so on \citep{efron2016computer}.

In Table \ref{statdic} are reported some descriptive statistics on the distribution of the Declared Tax Base for audited and not-audited units: the difference between two groups is really significant (t-test $185'295.75$ vs $92'834.80$; $p<0,001$).
\begin{table}
\centering
\begin{tabular}{lcrr}
\toprule
&  & Not-audited & Audited\\
\midrule
Mean & & $92'834.80$ & $185'295.75$   \\
Median & & $42'176$ & $73'362$   \\
Standard deviation & & $230'980.72$ & $387'869.98$    \\
Percentiles & 25 & $19'736$ & $30'532.75$    \\
 & 50 & $42'176$ & $73'362$    \\
  & 75 & $88'589$ & $169'942.50$    \\
\bottomrule
\end{tabular}
\caption{Descriptive statistics on the distribution of the Declared Tax Base for audited and not-audited units.}
\label{statdic}
\end{table}
Therefore, the tax-audit selection criteria is evidently skewed toward taxpayers with higher $\BID$, as proof that the sampling design depends on some individual characteristics with the tendency to select units with the higher potential to evade.
As a matter of fact, the selection may depend also on other variables and the more general picture is further investigated in Section \ref{Fit}. In the same section, the 2-steps Gradient Boosting and the Heckman model have been fitted to the same set of data in order to evaluate comparative performances.

\subsection{Model fitting and uncertainty assessment}
\label{Fit}
According to what has been explained in Section \ref{2steps}, two different gradient boosting have been subsequently fitted. The first one is a classification model aimed at estimating the inclusion probabilities while the second one is a weighted regressive model aimed at estimating the potential tax base. Both the models in the two steps have been fitted to the data and summarized in the software \texttt{R} using functions from the package \texttt{gbm} \citep{ridgeway2007generalized}. This function allows to decide in which way we want to evaluate the loss function at each iteration through the arguments \texttt{bag.fraction}, \texttt{train.fraction} and \texttt{cv.fold}: respectively bagging \citep{breiman1996bagging}, train-test split and complete cross-validation. As by default, our implementation relies on bagging (best compromise between speed and accuracy), with a \texttt{bag.fraction} value of $0.5$. 
The fit and validation of both the steps required about $1$ hour for a single run.

\paragraph{First step} The classification model of the first step has been fitted on the whole sub-sampled population of $64'207$ units, with the audited/not-audited variable as target ($45'489$ not-audited and $18'718$ audited). The final outcomes, properly re-scaled, are approximations $\left\lbrace\hat{\pi}_i\right\rbrace_{i=1}^N$ to the inclusion probabilities $\left\lbrace P(i\in s)\right\rbrace_{i=1}^N$ of each unit.
The Gradient Boosting fitting depends on some major tuning parameters among which the most relevant are: the number of iterations $n.iter$, the depth of each single tree $d$, the minimum number of observations in the final nodes of each tree and the learning parameter $\lambda$. They are not directly estimated by the model in the fitting process, but must be fixed before running the algorithm. An accurate choice of these parameters allow the GB to learn from the training data without incurring in over-fitting. The standard solution is to decide for a fixed grid of different combinations, fit the model for any possible of those and pick the one which returns the best performance in terms of some arbitrarily chosen metric. Unfortunately, it is not possible to try any possible combination and the refinement and extension of the grid must be chosen taking into account the computational time to fit and validate all the alternatives. Given the computational limitations introduced above, we had to limit ourselves to a very rough grid. The number of minimum observations in each final node has been fixed to the default value of $10$, while the tested $n.iter$, $d$ and $\lambda$ belonged to the following sets:
$$
n.iter\in\lbrace30, 40, \dots, 1000\rbrace,\qquad d\in \lbrace 2, 3\rbrace,\qquad \lambda\in \lbrace 0.01, 0.02,\dots,0.1\rbrace
$$
Evaluating the performances directly on the training set may be misleading since all machine learning techniques are very flexible and possibly affected by over-fitting \citep{friedman2001elements}. For this reason, the tuning parameters have been validated by splitting the sample into a \textit{training set} and a \textit{testing set}. The model is trained to estimate the probability of an audit using only information from the units in the training set and then its ability to recovered the outcome is validate on the testing set: the best set of parameters is the one that achieves the best score on the testing set.
In our application, $70\%$ of the units have been allocated to the training set while the remaining $30\%$ to the testing set (see Table \ref{traintest}). The metric chosen to evaluate the model performance in this step is the AUC score \citep{fawcett2006introduction}.

\begin{table}
\centering
\begin{tabular}{cccc}
  \toprule
 & Train set & Test set & \\ 
\midrule
Not-audited & $31'836$ & $13'653$ & $45'489$ \\
Audited & $13'064$ & $5'654$ & $18'718$\\
\midrule
& $44'900$ & $19'307$ & $64'207$ \\
\bottomrule
\end{tabular}
\caption{Training and testing set composition.}
\label{traintest}
\end{table}

The optimal choice was associated to the following set of parameters:
\begin{equation*}
\lbrace n.iter_{opt}=1000, \; d_{opt}=2, \;  \lambda_{opt}=0.1\rbrace \; .
\end{equation*}
and returned an AUC value of $0.8$ on the test data\footnote{ 
A greater value for $n.iter$, combined with lower values of $\lambda$, may provide even better results and this may be object of further investigation in future applications.}.

Together with the estimated inclusion probabilities, the \texttt{gbm} function returns also variables scores according to their importance in the fitting process (roughly speaking, the percentage of splits they determined). 
The most discriminating variables were the declared tax base ($\BID$), the activity branch, the dimension and the incomes of the firm, which are coherent with the results reported in Table \ref{statdic}. These may be considered proxy variables of the basic criteria adopted by the IRA to maximize the level of deterrence for the fiscal year 2011.

\begin{figure}[]
\centering
\includegraphics[width=0.8\linewidth]{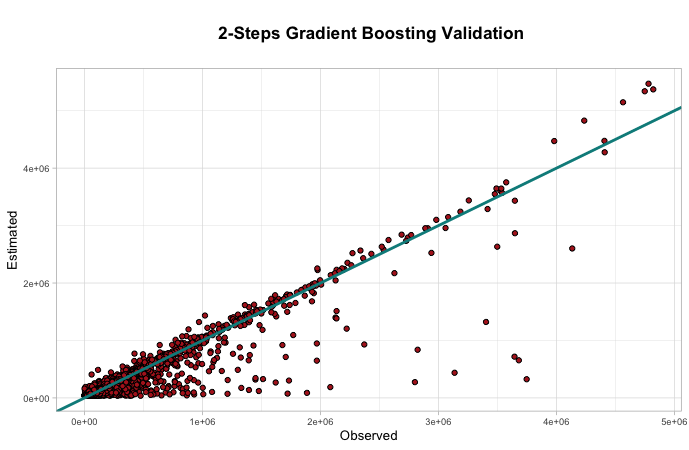}
\caption{Comparison of observed and predicted values of the $\BIT$ for the 2-steps GB, on the test set.}
\label{confronto}
\end{figure}

\paragraph{Second step} The regression model of the second step has been trained by weighting each unit by the inverse of its predicted inclusion probability $\hat{\pi_i}$. This can be done by utilizing the argument \texttt{weights} in the \texttt{gbm} function. In practice, when computing the loss function at each additional iteration of the GB, the error on each unit is weighted by $\nu_i$ as in equation \ref{pesi}: 
\begin{equation*}
\nu_i = \frac{1}{\hat{\pi_i}},\quad\forall\,i\in s.
\end{equation*}
Given that, the same procedure of the first step is adopted also for the validation of the second Gradient Boosting, which involved only the $18'718$ audited units (units in $s$). The sample is split in a train-set ($70\%$ of the units, $13'064$) and a test-set ($30\%$ of the units, $5'654$) as it is shown in Table \ref{traintest}. Since this step consisted of a regression problem, the optimal parameters have been chosen according to the $R^2$ index.
The best value obtained for the $R^2$ on the testing set is $0.83$, with tuning parameters:

\begin{equation*}
\lbrace n.iter_{opt}=380, \; depth_{opt}=2, \;  \lambda_{opt} = 0.1\rbrace \; 
\end{equation*}
Figure \ref{confronto} compares predicted and observed values on the test set: in case of perfect prediction all the points would be aligned on the bisector. We can notice how the most and the largest of the errors are related to the underestimation of the true potential tax base (lower section of the plot).

\begin{figure}[]
\centering
\includegraphics[width=0.8\linewidth]{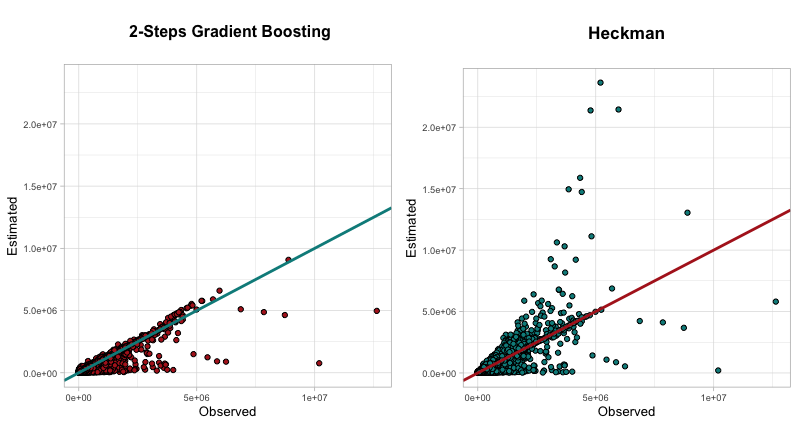}
\caption{Comparison of observed and predicted values of the $\BIT$ for the 2-steps GB (left) and the Heckman model (right), on all the audited units.}
\label{confronto2}
\end{figure}
Figure \ref{confronto2} visually compares the predictive performances of the 2-steps GB and the Heckman model on the whole set of audited units. It is worth mentioning that the Heckman model has been fitted on the log-transform of the outcome variable, so that the linear predictor could respect its natural domain. Predictions have been after back-transformed on the linear scale for performance evaluation. 
We can notice how the two models show an almost identical underestimation pattern, but the Heckman model also presents evident overestimation issues that are not present in the 2-steps GB. This could be caused by violation of the linearity assumption, which is pushing the Heckman predictions up for large values of the covariates.

Interval estimates for either the training set and the testing set have been produced using the technique in Section \ref{Unc} only on the second step of the procedure. The number of bootstrap samples from the training set has been fixed to $B=100$, each with the same size $n_B$ equal to the original training set size $13'064$. Therefore, at the end of the procedure, $B=100$ set of predictions $\left\lbrace\hat{Y}_{i,j}\right\rbrace_{j=1}^B$ are obtained for each observation $Y_i,\; i\in s$. The resulting intervals, obtained by computing the $5\%$ and $95\%$ quantiles of the empirical distributions of the bootstrapped predictions $\left\lbrace\hat{Y}_{i,}\right\rbrace_{i\in s}$, contained the true values $\left\lbrace Y_i\right\rbrace_{i\in s}$ only in the $30\%$ of the cases. The coverage is not satisfying, but it is important to highlight how the outcome predictions $\hat{Y}_{i,}$ are approximating the conditional expected value $E[Y_i|X_i]$. Therefore, the algorithm is bootstrapping the distribution of the prediction to $E[Y_i|X_i]$, which is by definition way less variable than the point observation $Y_i$ corresponding to the set of covariates $X_i$. Currently, there is not a unified framework for the production of prediction intervals in the Machine Learning framework. Other recently proposed techniques which focus on the uncertainty of the point observation are not discussed in this paper. They are based on bootstrapping the prediction error \citep{coulston2016approximating} or on building predictive models for the resulting predictions error \citep{shrestha2006machine}.
The good thing about the obtained interval is that the coverage is $30\%$ for both the training and the test set, reassuring us about the risk of over-fitting on the training data.
Furthermore, a greater value for $B$ is kindly suggested to improve the approximation. However, with only $B=100$ bootstrap sample parameters the algorithm took approximately $18$ hours to provide the final sets of predictions on the considered hardware. In general, it is not an very important issue since the computational time is linear in $B$ and can be drastically reduced using a better performing processor and parallelizing the procedure on a reasonable number of cores.

Finally, the predictions on the test set are compared to the ones produced by the standard Heckman model according to the same split. While the aggregate estimates of the total BIT result to be very close to each other, it is possible to notice differences in term of individual estimation. In particular, the $R^2$ obtained by the estimates from the Heckman model is equal to  $0.66$, which is sensibly lower than the one achieved by our new approach. Results are summarized in Table \ref{resume}.

\begin{table} \footnotesize
\centering
\begin{tabular}{lccc}
  \toprule
 & \textbf{Gradient Boosting} & \qquad \textbf{Heckman} & \textbf{Observed}\\
  \midrule
$\BIT$ & $1'292'440'659$ &\qquad  $1'230'785'551$ &\qquad $1'315'864'216$\\ 
$R^2$ & $0.828$ & \qquad $0.657$ & \qquad  \\ 
\bottomrule
\end{tabular}
\caption{Performances on the test set of the 2-steps Gradient Boosting and the Heckman model.}
\label{resume}
\end{table}

It seems that the Heckman Model is able to capture the general behaviour, but lacks of flexibility in order to get individual values. Linearity is a too strong restrain for such a complex problem, and the Gradient Boosting is not limited by this.

\subsection{An estimate of the propensity to evade VAT}
\label{PredProp}
Finally, the 2-steps Gradient Boosting approach has been used to produce predictions for all the units in the under-sampled population, including those whose BIT is unknown (not-audited taxpayers). Given the stratified structure of the sample, we trust that these estimates reflect properly the behaviour in the general population, with reasonable error margins. Such errors are not thoroughly discussed since the final aim of this section is not to actually provide an estimate for the total Italian VAT Gap, but just to show the potentiality of the application and the full potential that could be achieved by applying it to the whole set of data.

On the sample actually analyzed (see Table \ref{subsam}), it has been predicted a VAT \textit{gap} turnover $\hat{\BIND}=\hat{\BIT}-\BID$ of about $3.36Bln$ of euro ($3'360'930'741$\euro). This result is very close to the value returned by the Heckman model on the same sub-sample, which is of about $3.26Bln$ of euro ($3'260'653'691$\euro).
The credibility interval $[3.029Bln\,\text{\euro}, 3.555Bln\,\text{\euro}]$ has been obtained using the approach introduced in Section \ref{Unc}. It contains the value estimated by the Heckman model, highlighting again an inner coherency between the two techniques.

Furthermore, these predictions allowed the computation of a synthetic measure $p$ of \textit{VAT evasion propensity}, defined on the line of the evasion intensity used in \cite{braiotta2015tax}. This is defined, for each individual, as the ratio of the undeclared tax base and the potential tax base:
\begin{equation*}
\hat{p}_i = \frac{\hat{\BIND}_i}{\hat{\BIT}_i},\qquad i = 1,\dots, N.
\end{equation*}
Consequently, the general propensity to evade is estimated as:
\begin{equation*} 
\hat{p} = \frac{\sum_{i=1}^N \hat{\BIND}_i}{\sum_{i=1}^N \hat{\BIT}_i}
\end{equation*}
A low value of this ratio amounts to a compliant behaviour, and viceversa.

The observed VAT evasion propensity on all the audited taxpayers is of $23.09\%$. 
The estimated propensity using the fitted values of the 2-steps GB and the Heckman models on the same units are respectively of $22.12\%$ and $25.89\%$, with the former providing a result way closer to reality than the latter. 
Using the predictions on the entire sample of taxpayers (both audited and non audited) we get a value of $30.40\%$ for the 2-steps Gradient Boosting, slightly larger than the $29.77\%$ obtained with the Heckman model. 

Apparently, according to the results of both models, the Italian Revenue Agency is not selecting the individuals with the larger propensity to evade. Nevertheless, the observed average evaded by the audited taxpayers amounts to $55'637.09$, while the average evaded in the whole sample estimated by the GB is equal to $52'345.24$. This means that the Italian Revenue Agency sampling scheme favors the inclusion of individuals who evade the most in absolute terms, but not in relative terms: get the big evaders and recover as much gap as possible.
\begin{table} [] \scriptsize
\centering
\begin{tabular}{lcccccccccc} 
\toprule
&  & \multicolumn{3}{c}{\textbf{Observed}} & \multicolumn{3}{c}{\textbf{Gradient Boosting}} & 
\multicolumn{3}{c}{\textbf{Heckman}}\\
\textbf{Age} & \textbf{size}  & $ \mathbf{\underset{Bln}{BIT}}$ & $ \mathbf{\underset{Bln}{BIND}}$  & $\mathbf{Prop}$ & $\mathbf{\underset{Bln}{BIT}}$ & $ \mathbf{\underset{Bln}{BIND}}$  & $\mathbf{Prop}$ & $ \mathbf{\underset{Bln}{BIT}}$ & $ \mathbf{\underset{Bln}{BIND}}$  & $\mathbf{Prop}$\\
\midrule
\textbf{$[18-25)$} & $270$  & $0.06$ & $0.02$ & $25.24\%$  & $0.06$ & $0.01$ & $23.92\%$ &
$0.07$ & $0.02$ & $28.55\%$\\
\textbf{$[25-45)$} & $7876$  & $1.71$ & $0.43$ & $24.91\%$ & $1.69$ & $0.41$ & $23.97\%$  &
$1.76$ & $0.48$ & $26.98\%$\ \\
\textbf{$[45-65)$} & $9275$   & $2.35$ & $0.52$ & $22.25\%$ & $2.32$ & $0.49$ & $21.28\%$  &
$2.42$ & $0.59$ & $24.71\%$\ \\
\textbf{over $65$ } & $1297$  & $0.39$ & $0.08$ & $19.84\%$ & $0.38$ & $0.07$ & $18.69\%$   &
$0.43$ & $0.12$ & $27.67\%$\ \\
\midrule
\textbf{Total} & $18718$  & $4.51$ & $1.05$ & $23.09\%$ & $4.45$ & $0.98$ & $22.12\%$  &
$4.68$ & $1.21$ & $25.89\%$\ \\
\bottomrule
\end{tabular}
\caption{Observed and estimated propensity to evade by age classes on the audited taxpayers.}
\label{gendAgeAud}
\end{table}
In addition, estimated values from both models have been used to compute propensity of taxpayers grouped according to some of the observed covariates. These may be used to identify classes of individuals with high VAT evasion \textit{propensity} and may be of help in the selection procedure of future fiscal audit.
The propensity related to a specific class of individuals $c$ is straightforwardly estimated as:
\begin{equation*} 
\hat{p}_{c} = \frac{\sum_{i \in c} \hat{\BIND}_i}{\sum_{i \in c} \hat{BIT}_i}
\end{equation*}

\begin{table} [] \scriptsize
\centering
\begin{tabular}{lccccccc} 
\toprule
&  & \multicolumn{3}{c}{\textbf{Gradient Boosting}} & \multicolumn{3}{c}{\textbf{Heckman}}\\
\textbf{Age} & \textbf{size}  & $ \mathbf{\underset{Bln}{BIT}}$ & $ \mathbf{\underset{Bln}{BIND}}$  & $\mathbf{Prop}$ & $\mathbf{\underset{Bln}{BIT}}$ & $ \mathbf{\underset{Bln}{BIND}}$  & $\mathbf{Prop}$\\
\midrule
\textbf{$[18-25)$} & $976$  & $0.13$ & $0.05$ & $39.07\%$  & $0.13$ & $0.04$ & $36.71\%$  \\
\textbf{$[25-45)$} & $28250$  & $4.26$ & $1.45$ & $34.10\%$ & $4.11$ & $1.30$ & $31.61\%$ \\
\textbf{$[45-65)$} & $30496$   & $5.64$ & $1.60$ & $28.45\%$ & $5.64$ & $1.60$ & $28.37\%$\\
\textbf{over $65$ } & $4485$  & $1.02$ & $0.25$ & $24.65\%$ & $1.08$ & $0.32$ & $29.28\%$ \\
\midrule
\textbf{Total} & $64207$  & $11.05$ & $3.36$ & $30.40\%$ & $10.95$ & $3.26$ & $29.77\%$\\
\bottomrule
\end{tabular}
\caption{Estimated propensity to evade by age classes in the under-sampled population.}
\label{gendAge}
\end{table}
Unfortunately, description of the results must be limited due to the confidential nature of the data. In order to provide at least one example, we show the results related to the propensity by age. Age has been binned in four classes and the estimated propensity to evade is reported for all the classes. 
Considering only the audited taxpayers, the 2-steps GB returned results coherent with the observed values, while the Heckman model largely over-estimated the propensity and did not recover the true age trend (see Table \ref{gendAgeAud}). 
If we consider the whole sample, both the models reported decreasing propensities by age, but with some difference in terms of general behaviour. Indeed, the 2-steps Gradient Boosting emphasizes differences between classes: looking at Table \ref{gendAge}, it is possible to see how the $7$ points gap between the youngest and the oldest class estimated by the Heckman model becomes a $15$ points gap using the 2-steps GB.
Given the general better performances produced by the latter on the set of audited taxpayers, it is trustworthy that this greater variability is actually present in the population but the Heckman model is not able to catch it.

\section{Concluding remarks}
\addcontentsline{toc}{section}{Conclusions}
In this work, we present a non-parametric approach for the estimation of the VAT gap based on the Gradient Boosting algorithm, a machine learning technique for regression and classification problems. 
This new approach looks to be preferable and more suited to deal with this kind of data because it is distribution free, it can manage any kind of variable and it is not sensible to multicollinearity issues.
Furthermore, machine learning algorithms usually provide good performances also in high dimensional settings, and allow to exploit all the information contained in large sets of data.
The selection bias is accounted for using a different approach with respect to the one currently used by the Italian Revenue agency (Heckman model); the improved ability of our solution in retrieving information from a biased training set and producing good out-of-sample predictions is successfully verified in Appendix \ref{Toy}. The applicability of this approach extends to any \textit{learner} whose fit relies on the minimization of a loss function and, while our solution may look less sophisticated and rely on somewhat stronger assumptions on the origin of the bias, we think that it is consistent with the nature of the data and its actual origin.

In practice, the aggregate estimates of the VAT gap obtained through the two approaches are very similar; however, the Gradient Boosting based model produced sensibly more accurate estimates of the individual undeclared tax bases, catching the variability associated to observed variables as it is desirable. The Heckman model, on the contrary, seems to flatten out individual differences.

Accurate predictions at the individual level allowed the computation of trustworthy evasion propensity scores for all the taxpayers, detecting units and groups of units who are likely to hide a larger part of their incomes (in proportion). Unfortunately, a detailed description of the results referred to this index cannot be disclosed because of privacy issues.
At the aggregate level, the 2-steps Gradient Boosting estimates highlight how the audit selection process performed by the Italian Revenue Agency does not favor the selection of the \textit{less-compliant} individuals (those with a larger evasion propensity), but favors the selection of taxpayers with the largest $\BIND$ in absolute terms. Given the limited amount of resources and the consequent possibility to check only a small portion of the whole population, this choice sounds meaningful. However, it consistently neglects the (potentially very large) portion of population of small-evaders that do not report the most of their incomes.

The possible further developments of this kind of approach are various and promising. For instance, the analysis exposed in this work has been performed only on a small subset of all the available observations because of hardware limitations. We are confident that way better results may be achieved by analyzing the whole set of data. Moreover, further investigation of methods for the building of more reliable interval estimates is of main interest, being it one of the main drawbacks of the methodology.
Last but not least, improving the computational power would allow the application of a complete k-fold cross-validation for the tuning of the parameters, even on finer grids, and to apply more expensive but efficient techniques such as \textit{Neural Networks}.

\appendix

\section{Toy example: houses to rent}
\label{Toy}
We want to verify comparative performances of the Heckman model and our proposed 2-steps approach on a set of data in which we artificially induce a \textit{selection bias}.
We consider a dataset downloaded from Kaggle\footnote{\url{https://www.kaggle.com}} containing information about houses to rent in different cities in Brazil \citep{toyExamp}. It contains $10'962$ records, each with thirteen variables: two about the location of the house (city and area), six house-specific features (number of rooms, bathrooms\dots), and five economic amounts (homeowners association tax, rent amount, property tax\dots). 
We will consider the property tax, that is to say the annual tax the landlord should pay on the considered property, as our outcome variable of interest. It has been converted to the log-scale so that the Heckman linear predictor could respect its natural domain. The same log-transform has been considered also for the 2-steps GB in order to guarantee a fair comparison of the two methods.

\begin{figure}
    \centering
    \includegraphics[width=\textwidth]{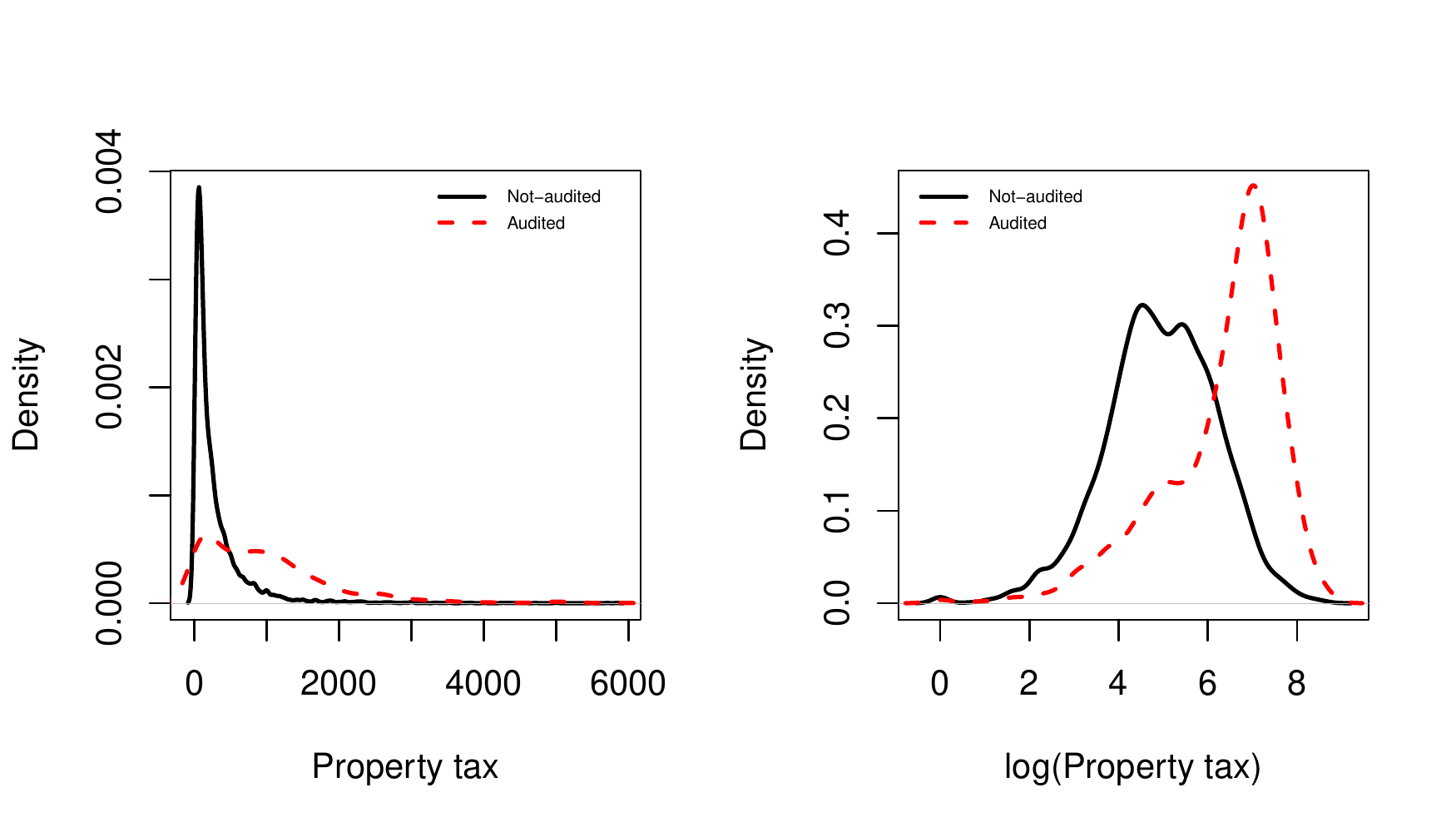}
    \caption{Distribution of the outcome variable in the "audited" and "not-audited" groups on the linear (left) and log (right) scales.}
    \label{biasBraz}
\end{figure}
A preliminary exploratory data analysis and a brief data cleaning step is required in order to favor proper fitting of both the models. The variable \textit{area} has been excluded from the analysis because it presents too many modalities (exactly $514$), of which many only have $1$ or $2$ observations (the maximization routine of the Heckman model would struggle in detecting all those parameters). The variable \textit{floor} is not available in the $23\%$ of the records and its value has been imputed using the complete average (any other imputation method, based on standard regression or regression trees, would favor one of the methodologies we are comparing). The \textit{property tax} is not available for $1596$ observations and, being this our outcome variable, such records have been omitted from the analysis. Finally, all the economic features present some very extreme outliers, which can be probably attributed to recording errors. In order to detect in an automatic way such values, all these variables have been converted to the log-scale and all instances trespassing the $q_{0.75}+1.5\times \text{IQR}$ threshold on this scale ($17$ records) have been deleted\footnote{$\text{IQR}$ is the \textit{interquartile range}: $q_{0.75}-q_{0.25}$}.

The resulting version of the dataset counts $9079$ records. Now, we need to select a subsample of observations to play the role of the \textit{audited taxpayers} (both covariates and outcome known), while the remaining will be the \textit{not-audited taxpayers} and will be used to test our predictive abilities (only covariates known). This selection must be performed in a way as similar as possible to how the Italian Revenue Agency selects the taxpayer to audit, i.e. by trying to maximize the selection of units with the highest outcome according to the available information. For instance, we may assume the \textit{Homeowner Association Tax} (HOA) as known and consider it a good proxy for the \textit{property tax} (PT). This association is confirmed by our data, which show a correlation of $\approx 0.55$ on a linear scale and of $\approx 0.67$ on the log-scale. We can then decide a cut-off point, let's say the $90$th percentile of the HOA, and consider as audited a small sample of units selected at random ($5\%$ of the total) plus all the units with HOA greater than the selected cut-off point. We end up with two groups, respectively composed of $1'304$ (audited) and $7'775$ (not audited) records. Figure \ref{biasBraz} shows the distribution of the outcome in the two groups and highlights the selection bias induced on this by our sampling scheme.
\begin{table}[t]
    \centering
    \begin{tabular}{l|cccc}
    \toprule
     & \multicolumn{2}{c}{Not-audited} & \multicolumn{2}{c}{Audited test set}\\
    \midrule
    Model     &  MSE & $R^2$ & MSE & $R^2$\\
    \midrule
    Heckman     & 9.696  & 6.32 & 0.738 &  0.4 \\
    2-steps GB     & 0.762 &  0.422 & 0.716 & 0.388 \\
    \bottomrule
    \end{tabular}
    \caption{Error metrics of the two considered models on the not-audited set and the audited test set units in terms of MSE and $R^2$.}
    \label{resTab}
\end{table}
In order to test the performance of the final predictive models also on the "audited" units, only the $70\%$ of those are used in the training process while the remaining $30\%$ are kept out as testing set.

The Heckman model is fitted to the data through the function \texttt{heckit} from the \texttt{sampleSelection} package available on the CRAN, which provides estimation routine for a variety of tobit-like models.
The \textit{2-steps Gradient Boosting} is fitted through the function \texttt{gbm} from the \texttt{gbm} package available on the CRAN. $5$-folds cross-validation (argument \texttt{cv.folds} of the \texttt{gbm} function) has been performed on the same grid of the original application introduced in Section \ref{App}.

The final prediction accuracy has been measured in terms of \textit{Mean Squared Error} (MSE) on the test portion of the "audited" units and on all the "not-audited" units, which is the group on which we are trying to achieve the best accuracy.
In order to re-scale by the \textit{prediction difficulty} of each set, also the $R^2$ (Relative Mean Squared Error) is provided. It is computed dividing the MSE by the variance of the outcome in the considered set.
Results are summarized in Table \ref{resTab}.

\begin{figure}
    \centering
    \includegraphics[width=\textwidth]{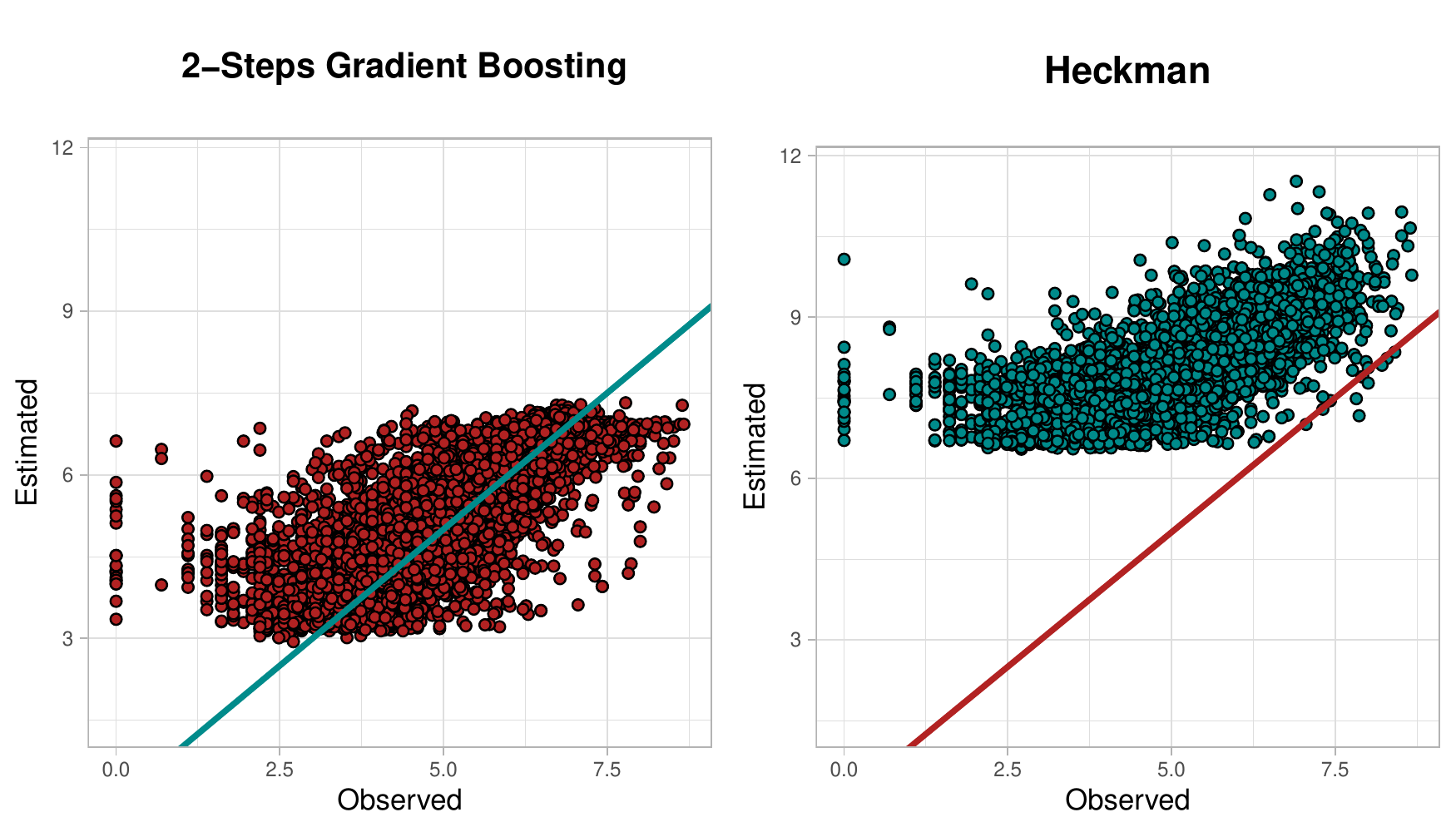}
    \caption{Comparison of observed and predicted values for the 2-steps GB (left) and the Heckman model (right) on the not-audited units.}
    \label{PredBraz}
\end{figure}

We can clearly notice how the Gradient Boosting largely outperforms the Heckman model, producing a sensibly lower errors especially on the set of not-audited units. 
A graphical comparison of the prediction accuracy on the non-audited units is provided in Figure \ref{PredBraz}.

We can then conclude that the non-linearity and model free-ness of the Gradient Boosting, combined with proper weighting to account for the non-random selection of the training units, produce a sensible improvement whenever sample selection is performed according to (unknown) criteria based on the available covariates.

\bibliographystyle{plainnat}  
\bibliography{VAT}  

\begin{thebibliography}{39}
\providecommand{\natexlab}[1]{#1}
\providecommand{\url}[1]{\texttt{#1}}
\expandafter\ifx\csname urlstyle\endcsname\relax
  \providecommand{\doi}[1]{doi: #1}\else
  \providecommand{\doi}{doi: \begingroup \urlstyle{rm}\Url}\fi

\bibitem[Alho(1990)]{alho1990adjusting}
Juha~M Alho.
\newblock Adjusting for nonresponse bias using logistic regression.
\newblock \emph{Biometrika}, 77\penalty0 (3):\penalty0 617--624, 1990.

\bibitem[Amemiya(1986)]{amemiya1986advanced}
Takeshi Amemiya.
\newblock Advanced econometrics (1985 ed.), 1986.

\bibitem[Austin and Stuart(2015)]{austin2015moving}
Peter~C Austin and Elizabeth~A Stuart.
\newblock Moving towards best practice when using inverse probability of
  treatment weighting (iptw) using the propensity score to estimate causal
  treatment effects in observational studies.
\newblock \emph{Statistics in medicine}, 34\penalty0 (28):\penalty0 3661--3679,
  2015.

\bibitem[Bethlehem(1988)]{bethlehem1988reduction}
Jelke~G Bethlehem.
\newblock Reduction of nonresponse bias through regression estimation.
\newblock \emph{Journal of Official Statistics}, 4\penalty0 (3):\penalty0 251,
  1988.

\bibitem[Braiotta et~al.(2015)Braiotta, Carfora, Pansini, and
  Pisani]{braiotta2015tax}
Alessandra Braiotta, Alfonso Carfora, Rosaria~Vega Pansini, and Stefano Pisani.
\newblock Tax gap and redistributive aspects across italy.
\newblock \emph{Italian Revenue Agency Discussion Topics}, 2:\penalty0 1--27,
  2015.

\bibitem[Breiman(1996)]{breiman1996bagging}
Leo Breiman.
\newblock Bagging predictors.
\newblock \emph{Machine learning}, 24\penalty0 (2):\penalty0 123--140, 1996.

\bibitem[Breiman et~al.(1984)Breiman, Friedman, Stone, and
  Olshen]{breiman1984classification}
Leo Breiman, Jerome Friedman, Charles~J Stone, and Richard~A Olshen.
\newblock \emph{Classification and regression trees}.
\newblock CRC press, 1984.

\bibitem[Chawla(2009)]{chawla2009data}
Nitesh~V Chawla.
\newblock Data mining for imbalanced datasets: An overview.
\newblock In \emph{Data mining and knowledge discovery handbook}, pages
  875--886. Springer, 2009.

\bibitem[Coulston et~al.(2016)Coulston, Blinn, Thomas, and
  Wynne]{coulston2016approximating}
John~W Coulston, Christine~E Blinn, Valerie~A Thomas, and Randolph~H Wynne.
\newblock Approximating prediction uncertainty for random forest regression
  models.
\newblock \emph{Photogrammetric Engineering \& Remote Sensing}, 82\penalty0
  (3):\penalty0 189--197, 2016.

\bibitem[Dangerfield and Morris(1992)]{dangerfield1992top}
Byron~J Dangerfield and John~S Morris.
\newblock Top-down or bottom-up: Aggregate versus disaggregate extrapolations.
\newblock \emph{International journal of forecasting}, 8\penalty0 (2):\penalty0
  233--241, 1992.

\bibitem[D’Agosto et~al.(2016)D’Agosto, Marigliani, and
  Pisani]{d2016general}
E~D’Agosto, M~Marigliani, and S~Pisani.
\newblock A general framework for measuring vat compliance in italy.
\newblock \emph{Argomenti di discussione dell’Agenzia delle Entrate},
  \penalty0 (2), 2016.

\bibitem[D’Agosto et~al.(2014)D’Agosto, Marigliani, and
  Pisani]{d2014asymmetries}
Elena D’Agosto, Massimiliano Marigliani, and Stefano Pisani.
\newblock Asymmetries in the territorial vat gap.
\newblock \emph{Argomenti di Discussione of Italian Revenue Agency}, 2, 2014.

\bibitem[Efron and Hastie(2016)]{efron2016computer}
Bradley Efron and Trevor Hastie.
\newblock \emph{Computer age statistical inference}, volume~5.
\newblock Cambridge University Press, 2016.

\bibitem[Efron et~al.(2003)]{efron2003second}
Bradley Efron et~al.
\newblock Second thoughts on the bootstrap.
\newblock \emph{Statistical Science}, 18\penalty0 (2):\penalty0 135--140, 2003.

\bibitem[Fawcett(2006)]{fawcett2006introduction}
Tom Fawcett.
\newblock An introduction to roc analysis.
\newblock \emph{Pattern recognition letters}, 27\penalty0 (8):\penalty0
  861--874, 2006.

\bibitem[Friedman et~al.(2001)Friedman, Hastie, and
  Tibshirani]{friedman2001elements}
Jerome Friedman, Trevor Hastie, and Robert Tibshirani.
\newblock \emph{The elements of statistical learning}, volume~1.
\newblock Springer series in statistics New York, 2001.

\bibitem[Friedman(2001)]{friedman2001greedy}
Jerome~H Friedman.
\newblock Greedy function approximation: a gradient boosting machine.
\newblock \emph{Annals of statistics}, pages 1189--1232, 2001.

\bibitem[Group(2018)]{taxpap_0073}
FISCALIS Tax Gap~Project Group.
\newblock {The concept of tax gaps - Corporate Income Tax Gap Estimation
  Methodologies}.
\newblock Taxation Papers~73, Directorate General Taxation and Customs Union,
  European Commission, November 2018.
\newblock URL \url{https://ideas.repec.org/p/tax/taxpap/0073.html}.

\bibitem[{He} and {Garcia}(2009)]{he2009Imb}
H.~{He} and E.~A. {Garcia}.
\newblock Learning from imbalanced data.
\newblock \emph{IEEE Transactions on Knowledge and Data Engineering},
  21\penalty0 (9):\penalty0 1263--1284, 2009.

\bibitem[Heckman(1976)]{heckman1976common}
James~J Heckman.
\newblock The common structure of statistical models of truncation, sample
  selection and limited dependent variables and a simple estimator for such
  models.
\newblock In \emph{Annals of Economic and Social Measurement, Volume 5, number
  4}, pages 475--492. NBER, 1976.

\bibitem[Heckman(1979)]{heckman1979sample}
James~J Heckman.
\newblock Sample selection bias as a specification error.
\newblock \emph{Econometrica: Journal of the econometric society}, pages
  153--161, 1979.

\bibitem[Heskes(1997)]{heskes1997practical}
Tom Heskes.
\newblock Practical confidence and prediction intervals.
\newblock In \emph{Advances in neural information processing systems}, pages
  176--182, 1997.

\bibitem[Hirano et~al.(2003)Hirano, Imbens, and Ridder]{hirano2003efficient}
Keisuke Hirano, Guido~W Imbens, and Geert Ridder.
\newblock Efficient estimation of average treatment effects using the estimated
  propensity score.
\newblock \emph{Econometrica}, 71\penalty0 (4):\penalty0 1161--1189, 2003.

\bibitem[Kumar et~al.(2015)Kumar, Rao, et~al.]{kumar2015minimising}
Sudhanshu Kumar, R~Kavita Rao, et~al.
\newblock Minimising selection failure and measuring tax gap: An empirical
  model.
\newblock Technical report, 2015.

\bibitem[More(2016)]{more2016survey}
Ajinkya More.
\newblock Survey of resampling techniques for improving classification
  performance in unbalanced datasets.
\newblock \emph{arXiv preprint arXiv:1608.06048}, 2016.

\bibitem[OECD(2017)]{OECD2017}
OECD.
\newblock \emph{Tax Administration 2017}.
\newblock 2017.
\newblock URL
  \url{https://www.oecd-ilibrary.org/content/publication/tax_admin-2017-en}.

\bibitem[Pisani and Pansini(2017)]{pisaniBott}
Stefano Pisani and Rosaria~vega Pansini.
\newblock Bottom-up estimates of tax gap by the italian revenue agency, 2017.
\newblock URL
  \url{http://docplayer.net/139929090-Bottom-up-estimates-of-tax-gap-by-the-italian-revenue-agency.html}.

\bibitem[Puhani(2000)]{puhani2000heckman}
Patrick Puhani.
\newblock The heckman correction for sample selection and its critique.
\newblock \emph{Journal of economic surveys}, 14\penalty0 (1):\penalty0 53--68,
  2000.

\bibitem[Ridgeway(2007)]{ridgeway2007generalized}
Greg Ridgeway.
\newblock Generalized boosted models: A guide to the gbm package.
\newblock \emph{Update}, 1\penalty0 (1):\penalty0 2007, 2007.

\bibitem[Rubens(2020)]{toyExamp}
Junior Rubens.
\newblock {brazilian\_ houses\_ to\_ rent [Version 2]}, 2020.
\newblock URL
  \url{https://www.kaggle.com/rubenssjr/brasilian-houses-to-rent#houses_to_rent_v2.csv}.
\newblock {Data retrieved from \href{https://www.kaggle.com/}{Kaggle}}.

\bibitem[Santoro(2010)]{santoro2010evasione}
Alessandro Santoro.
\newblock \emph{L'evasione fiscale:[quanto, come e perch{\'e}]}.
\newblock Il mulino, 2010.

\bibitem[S{\"a}rndal and Lundstr{\"o}m(2005)]{sarndal2005estimation}
Carl-Erik S{\"a}rndal and Sixten Lundstr{\"o}m.
\newblock \emph{Estimation in surveys with nonresponse}.
\newblock John Wiley \& Sons, 2005.

\bibitem[S{\"a}rndal et~al.(2003)S{\"a}rndal, Swensson, and
  Wretman]{sarndal2003model}
Carl-Erik S{\"a}rndal, Bengt Swensson, and Jan Wretman.
\newblock \emph{Model assisted survey sampling}.
\newblock Springer Science \& Business Media, 2003.

\bibitem[Schapire(1990)]{schapire1990strength}
Robert~E Schapire.
\newblock The strength of weak learnability.
\newblock \emph{Machine learning}, 5\penalty0 (2):\penalty0 197--227, 1990.

\bibitem[Shrestha and Solomatine(2006)]{shrestha2006machine}
Durga~L Shrestha and Dimitri~P Solomatine.
\newblock Machine learning approaches for estimation of prediction interval for
  the model output.
\newblock \emph{Neural Networks}, 19\penalty0 (2):\penalty0 225--235, 2006.

\bibitem[Tobin(1952)]{tobin1952survey}
James Tobin.
\newblock A survey of the theory of rationing.
\newblock \emph{Econometrica: Journal of the Econometric Society}, pages
  521--553, 1952.

\bibitem[Wickham and Wickham(2016)]{wickham2016package}
Hadley Wickham and Maintainer~Hadley Wickham.
\newblock Package ‘tidyr’, 2016.

\bibitem[Wickham et~al.(2015)Wickham, Francois, Henry, M{\"u}ller,
  et~al.]{wickham2015dplyr}
Hadley Wickham, Romain Francois, Lionel Henry, Kirill M{\"u}ller, et~al.
\newblock dplyr: A grammar of data manipulation.
\newblock \emph{R package version 0.4}, 3, 2015.

\bibitem[Zadrozny(2004)]{zadrozny2004learning}
Bianca Zadrozny.
\newblock Learning and evaluating classifiers under sample selection bias.
\newblock In \emph{Proceedings of the twenty-first international conference on
  Machine learning}, page 114. ACM, 2004.

\end{thebibliography}






\end{document}